\documentclass[authoryear,preprint,12pt]{elsarticle}

\journal{Icarus}

\begin{document}

\begin{frontmatter}

\title{A note on cement in asteroids}

\author{G. Bilalbegovi\' c}
\ead{goranka.bilalbegovic@gmail.com}
\address{Department of Physics, Faculty of Science, University of Zagreb, Bijeni\v cka 32, 10000 Zagreb, Croatia}

\begin{abstract}
Cement mineral tobermorite was formed in hydrothermal experiments on alternation of 
calcium-aluminum-rich inclusions (CAIs) in carbonaceous chondrite meteorites.
Unidentified bands at 14 $\mu$m  were measured for CAIs and the matrix of the Allende
meteorite sample, as well as for Hektor and Agamemnon asteroids. The presence of cement nanoparticles  may explain
the feature at 14 $\mu$m.

\end{abstract}


\end{frontmatter}

Cement is one of the most used materials on the Earth.   Concrete is made from cement paste, sand, and stone. Cement paste consists of cement powder and water. The main component of cement, calcium-silicate-hydrate, has a  granular complex structure  at nanoscales (see Fig. 1 in \cite{Allen2007}). Although concrete and cement we use are man-made materials, several cement crystals are naturally formed on the Earth. The structure of cement is sometimes described using minerals tobermorite and jennite, but many other natural terrestrial forms of calcium-silicate-hydrates exist (see Table 1 in \cite{Richardson2008}). 
For example, even tobermorite is not the unique crystal, but appears in three main forms, known as 9 \AA, 11 \AA, and 14 \AA\, tobermorites. These systems vary in the interlayer distance. It is found that more than 30 stable phases of calcium-silicate-hydrates form in hydrothermal environments \citep{Shaw2000}.
Two of these phases are 11 \AA\, tobermorite (Ca$_5$Si$_6$O$_{16}$(OH)$_2$ $\cdot$ 4 H$_2$O)  and xonotlite (Ca$_6$Si$_6$O$_{17}$(OH)$_2$).

A recent study of cement nanoparticles revealed that,  together with typical bands of amorphous silicates at 10 and 18 $\mu$m, these structures show specific infrared  (IR) features at 14 $\mu$m \citep{Bilalbegovic2014}. 
Figs.  2-4 in \cite{Bilalbegovic2014} show that bands of cement nanoparticles exist in near-, mid- and far-IR  spectral regions.   IR bands are  described  and features  with higher intensities are listed in Table  1 of \cite{Bilalbegovic2014}.
Measurements of IR spectra for the cement paste showed that the feature at  14 $\mu$m exists for a range of compositions and it
is generated by the Si-O-Si bending vibrations \citep{Garbev2007}. However, it is known that the frequency of these bending vibrations changes with silicate structure and polymerization.
It should be expected that exact positions of bands close to 14 $\mu$m vary with the composition and size of cement nanoparticles.

It is possible that cement forms in various environments in Space.
The Infrared Space Observatory (ISO) and the Spitzer telescope observed  unidentified IR features at 14 $\mu$m for several stars: HD 44179 (Red rectangle), HD 45677, and IRC+10420  \citep{Molster2002,Kemper2005}. Recent data for  the planetary nebula BD+303639
taken by the FORCAST instrument on board of the Sofia telescope have also shown the unidentified 14 $\mu$m band (Guzman-Ramirez, personal communications).
Table 1 compares positions of IR bands at 14 $\mu$m, which were calculated, measured, or observed for several objects on the Earth and in Space.

It is known that some meteorites contain calcium-aluminum-rich inclusions (CAIs) that are either pre-solar grains, or the first material condensed in the Solar nebula \citep{Krot2005}.  \cite{Nomura1998} carried out experiments on alternation of CAIs in carbonaceous chondrites to study aqueous alternation in parent asteroids.
They investigated some of the oldest phases in the  Solar System: gehlenite, spinel, and diopside,  as well as mixtures of gehlenite with SiO$_2$ and Al$_2$O$_3$. One of the products obtained in hydrothermal experiments from the mixture of gehlenite with  SiO$_2$  was tobermorite.  This mineral is one of Ca-Si hydrates and exists as natural cement on the Earth \citep{Richardson2008}. 
 \cite{Nomura1998}
 suggested that the formation of Ca-Si hydrates started when gehlenite was disolved by a fluid and Ca, Al, and Si atoms were released.  Tobermorite  crystallized  when sufficient numbers of Ca and Si atoms were present. \cite{Nomura1998} wrote that \cite{Shoji1974a,Shoji1974b} in hydrothermal experiments
 obtained another Ca-Si hydrate mineral xonotlite. The hydrothermal formation of xonotlite was later investigated in details \citep{Shaw2000}.
 Hydrous minerals produced in asteroids by the process of aqueous alternation could convert into anhydrous ones by thermal metamorphism.
 \cite{Nomura1998} found that tobermorite exists in a whisker form and converts to wollastonite with increasing temperature. 
 It is possible that both tobermorite and wollastonite exist in the same environment if thermal metamorphism is not complete.

The Allende meteorite is from the oxidized subgroup of the CV carbonaceous chondrites.
Allende was affected by aqueous alternation on the parent body \citep{Krot1998}.  It is known that both gehlenite \citep{Nomura1998}
and wollastonite \citep{Miyamoto1979,Krot1998}
are present in Allende. Therefore, it is possible that tobermorite and other calcium-silicate-hydrates
were also formed in the parent asteroid body and could be detected in Allende. 
CAIs are minor components of chondrite meteorites. In CV chondrites refractory inclusions (CAIs and AOAs  (amoeboid olivine aggregates) together)
make about 10 vol. 10\% \citep{Scott2005}. Therefore, the amount of cement expected to form in the processes suggested by \cite{Nomura1998} is not big.
Morlok and coworkers studied mid-IR spectra (between 2.5 $\mu$m and 16 $\mu$m)  of CAIs separated from Allende
\citep{Morlok2008}.  The sample analyzed was from the collection of the Natural History Museum in London and it was collected immediately after its fall. Therefore, this sample was not substantially changed by conditions on the Earth.  The sample measured was a fine-grained, submicron-sized powder. It was difficult to characterize the minerals. Therefore,  mixtures of several phases were analyzed. 
CaO i SiO$_2$ were detected.
Table 2 shows IR bands at 14 $\mu$m measured for materials in the sample of Allende
\citep{Morlok2008}. Morlok and coworkers suggested that a carrier of the  broad feature between 14.0 $\mu$m and 15.0 $\mu$m could be  spinel. 
For example, \cite{Chihara2000} studied the optical constants of three crystalline spinels (MgAl$_2$O$_4$) in the mid- and far-infrared spectral regions. Two samples were natural spinels (white and pink), whereas one was synthesized. The pink spinel included a small amount of Cr, and synthetic spinel was Al$_2$O$_3$ rich. Bands at 14 $\mu$m were measured for the pink spinel (14.7 $\mu$m in the absorption spectrum), the white spinel (14.8 $\mu$m in absorption), and the synthetic spinel (14.6 $\mu$m in absorption and 14.0 $\mu$m in reflection).
Morlok and coworkers  proposed that a position within the broad band probably depends on the spinel's  iron content. However, IR spectra of iron-bearing spinels were measured \citep{Cloutis2004,Jackson2014}. Cloutis and coworkers found that the best correlation exists between Fe$^{2+}$ content and wavelength positions at
0.46, 0.93, 2.8, 12.3, 16.2, and 17.5 $\mu$m. For Fe$^{3+}$ content the best correlation is at 0.93 $\mu$m. The authors did not mention any relevance of the 
(14-15) $\mu$m spectral region for a broad range of chemical compositions of iron-bearing spinels they studied \citep{Cloutis2004,Jackson2014}. IR features of cement can explain bands at 14 $\mu$m measured by Morlok and coworkers.

\cite{Emery2006} analyzed spectra of Trojan asteroids  624 Hektor, 911 Agamemnon, and 1172 Aneas, 
obtained by the IR spectrograph (IRS) on the Spitzer Space Telescope.  After comparisons of their data with available spectral libraries for meteorites and minerals, these authors concluded that small silicate grains suspended in a transparent matrix are the best candidates to reproduce mid-IR spectral shapes of asteroids they studied, such as the Si-O stretch fundamental at 10  $\mu$m and the Si-O bend fundamental at $\sim$18 to 25 $\mu$m.
However, comparisons with existing spectral databases for meteorites and minerals did not explain all details of asteroid spectra. A small, but visible, spectral feature between  14 $\mu$m and  15 $\mu$m was found for Hektor and Agamemnon.  Cement nanoparticles are good candidates for carriers of this feature. 

Hydrated minerals in asteroids are traced by absorption features at 0.7 and 3 $\mu$m.  Observations of Trojan asteroids in the visible 
\citep{Bendjoya2004} and near-IR  \citep{Cruikshank2001,Emery2003} spectral regions did not show these bands. Therefore, till now there is no evidence of hydrated minerals  on surfaces of Trojan asteroids and a route to the cement formation  on Hektor and Agamemnon based on altered CAIs is not supported by available observations.
The Tagish lake carbonaceous chondrite shows similar spectra as the D-type asteroids and it was proposed as their analog  \citep{Hiroi2001}.  
A recent analysis shows that possible parent bodies for Tagish Lake are very rare in the group of Jupiter Trojans \citep{Vernazza2013}.
The Tagish Lake meteorite is carbon rich and has high concentrations of presolar grains, as well as carbonate minerals. The presence of these ingredients  is consistent with the present understanding of compositions of the D-type asteroids. However, Trojans have redder spectral slopes and
the Tagish lake  meteorite is aqueously altered.  Several ideas have been put forward to explain the non-detection of hydrated minerals on Trojans.
\cite{Cruikshank2001} proposed that hydrated minerals on the surface of Hektor could be masked with low-albedo materials, such as carbon.  They also suggested that the strength of 3 $\mu$m  band decreases with heliocentric distance because of increasing abundance of carbon. 
Dehydration by micrometeorite bombardments of the surface of some D-type asteroids  was also proposed \citep{Hiroi2001}. Future high-resolution observations of Trojan asteroids, as well as a fall of a meteorite similar to the Jupiter Trojan D-type asteroids, would be able to settle this problem.

Experiments on alternation of CAIs  by \cite{Nomura1998} shown that cement forms in carbonaceous chondrites. 
 Ca atoms  are supplied by gehlenite in this hydrothermal scenario of cement formation. 
It is known that gehlenite (Al-rich melilite) is a typical mineral in CAIs \citep{Nomura1998,Morlok2008}.
\cite{Emery2006} 
used a modified standard thermal model to estimate the surface temperatures of Trojan asteroids they studied. Because of insufficient information on the physical
properties, they also varied several parameters to find the best fit to the measured spectral energy distribution and 
found that the Hector's surface temperature  is 179.4 K. 
\cite{Nomura1998} performed their experiments on alternation of CAIs in carbonaceous chondrites where tobermorite was formed at 473.15 K.
The presence of water ice and other ices in Trojan asteroids was studied \citep{Cruikshank2001,Yang2007,GuilbertLepoutre2014}.  Present understanding is that  Trojans have a large fraction of water ice covered by a refractory mantle \citep{Emery2015}. Yang and co-workers suggested that in the young Solar system short-lived radionuclides heated Trojans and their ices melted \citep{Yang2013}. A mechanical mixture of water, gehlenite and other primitive silicates,  as well as other soluble materials circulated. These conditions were on some asteroids favorable for the formation of cement nanoparticles.

Cement, as a specific silicate, should be included in spectral databases for future studies of asteroids and meteorites.

This work is supported by the HRZZ Research Project ``Stars and dust: structure, composition and interaction" and the QuantiXLie Center of Excellence.
I would like to thank Lizette Guzman-Ramirez for interesting discussions about the BD+303639 planetary nebula and IR spectra.
This research has made use of NASA's Astrophysics Data System Bibliographic Services.

\begin{center}
\begin{table*}
 \caption{IR bands at 14 $\mu$m for several cement structures and corresponding unidentified features  in astrophysical objects. Data are from:
 (a) \cite{Bilalbegovic2014}, (b) \cite{Garbev2007}, (c) \cite{Kemper2005}, (d) \cite{Molster2002}, (e) \cite{Emery2006}.}
\label{tab1}
 \begin{tabular}{l   l  l } 
 \hline 
 Source & State & IR at 14 $\mu$m \\  
 \hline \hline
 Ca$_4$Si$_4$O$_{14}$H$_4$ & nanoparticle, crystalline precursor & 14.44, 14.55 $^{(a)}$\\ 
 \hline
 Ca$_6$Si$_3$O$_{13}$H$_2$ & nanoparticle, amorphous & 14.10 $^{(a)}$\\
 \hline
 Ca$_{12}$Si$_6$O$_{26}$H$_4$ & nanoparticle, amorphous & 14.58 $^{(a)}$\\
 \hline
 Cement paste & compositions CaO/SiO$_2$ = 0.2 -1.5 & 14.88 - 14.97 $^{(b)}$\\
 \hline
 Red  Rectangle (Spitzer)  & unknown & 14.-16. $^{(c)}$\\ 
 \hline
 Red Rectangle (ISO) & unknown & 14.19 $^{(d)}$ \\ 
 \hline
 HD 45677  & unknown & 14.3 $^{(d)}$\\
 \hline
 IRC+10420  & unknown & 14.15 $^{(d)}$\\
 \hline
 Hektor, Agamemnon & unknown & 14.-15. $^{(e)}$ \\
 \hline
\end{tabular}
\end{table*}
\end{center}

\begin{center}
\begin{table*}
 \caption{IR bands at 14 $\mu$m for the matrix and CAIs of the Allende meteorite  sample \citep{Morlok2008}.}
\label{tab2}
 \begin{tabular}{l   l } 
 \hline 
 Object & Bands at 14 $\mu$m \\  
 \hline \hline
 Matrix & 14.29, 14.97\\ 
 \hline
 RIM 1 & 14.62, 14.88\\
 \hline
 RIM 2 & 14.62, 14.88\\
 \hline
 Interior 1 &  14.12, 14.39, 14.97\\
 \hline
 Interior 2 & 14.21, 14.88\\ 
 \hline
 Interior 4 &  14.88 \\ 
 \hline
 Interior (close RIM) & 14.97\\
 \hline
\end{tabular}
\end{table*}
\end{center}

\bibliographystyle{elsarticle-harv} 
\bibliography{jtrojans}

\end{document}